\documentclass{PoS}

\usepackage{siunitx}
\usepackage{upgreek}
\usepackage{subfig}

\newcommand{\mueee}{\mbox{$\upmu\,\rightarrow\,\text{eee}$}}
\newcommand{\muposeee}{\mbox{$\upmu^+\,\rightarrow\,\text{e}^+\text{e}^-\text{e}^+$}}

\newcommand{\muposeeenunu}{\mbox{$\upmu^+\,\rightarrow\,\text{e}^+\text{e}^-\text{e}^+\overline{\upnu}_\upmu\upnu_\text{e}$}}

\newcommand{\BR}{\text{BR}}

\DeclareSIUnit[]\muon{\ensuremath{\upmu}}

\title{Searching for Lepton Flavour Violation with the Mu3e Experiment}

\ShortTitle{Searching for Lepton Flavour Violation with Mu3e}

\author{\speaker{Ann-Kathrin Perrevoort} for the Mu3e Collaboration%%\thanks{A footnote may follow.}
  \\
        Physics Institute, Heidelberg University\\
        E-mail: \email{perrevoort@physi.uni-heidelberg.de}}

\abstract{The upcoming Mu3e experiment searches for the lepton flavour
  violating decay \muposeee\ with the aim of a final sensitivity of one signal
  decay in $\num{e16}$ observed muon decays, an improvement over the preceding
  SINDRUM experiment by four orders of magnitude. In the first phase, the
  experiment will be operated at an existing intense muon beam line at the
  Paul Scherrer Institute. With muon stopping rates of about
  $\num{e8}\si{\per\second}$, a single-event sensitivity of $\num{2e-15}$ can
  be achieved. For the ultimate sensitivity, a new high intensity muon beam
  line is required. \\
  In order to suppress background, the tracking detector is designed to
  measure low momentum electron and positron tracks with excellent precision
  by making use of very thin silicon pixel sensors. In addition, scintillating
  fibres and tiles provide precise timing information. \\
  Currently, the collaboration is finalizing the detector design and preparing
  for construction and commissioning.

}

\FullConference{The 19th International Workshop on Neutrinos from Accelerators-NUFACT2017\\
		25-30 September, 2017\\
		Uppsala University, Uppsala, Sweden}

\begin{document}

\section{Introduction}
\begin{figure}[tbp]
  \centering
  \subfloat[Neutrino mixing.]{\includegraphics[height=0.13\textheight,width=0.33\textwidth]{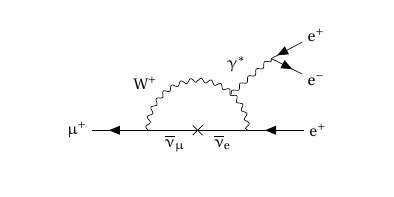}\label{FigMu3eMixing}}\quad
  \subfloat[Supersymmetric particles in a loop
    diagram.]{\includegraphics[height=0.13\textheight,width=0.33\textwidth]{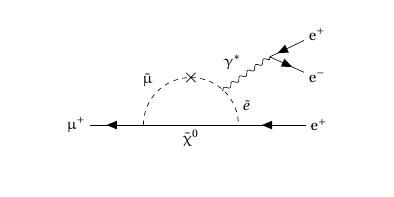}\label{FigMu3eSUSY}}\quad
  \subfloat[At tree level via a Z$^\prime$.]{\includegraphics[height=0.16\textheight,width=0.28\textwidth]{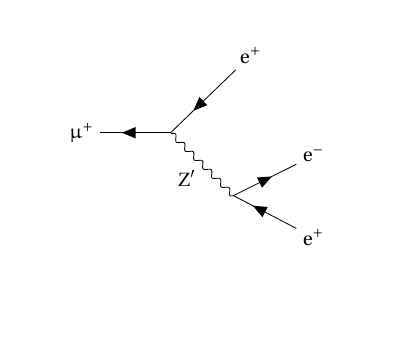}\label{FigMu3eZprime}} 
  \caption{The \mueee\ decay mediated via different processes. }
          \label{FigMu3eDecay}
\end{figure}

In the Standard Model of particle physics, lepton flavour is conserved. Yet
with the observation of neutrino oscillations, it became evident that nature
does not conserve lepton flavour, although an observation of lepton
flavour violation for charged leptons is still missing.
Extending the Standard Model to include neutrino mixing, lepton flavour
violating muon decays for example are mediated in loop-diagrams (see
figure~\ref{FigMu3eMixing}). With branching fractions of below
$\num{e-54}$, these decays are far beyond experimental reach. Hence,
any observation of charged lepton flavour violation would be a clear sign for
physics beyond the Standard Model.\\
One channel to search for such phenomena is \muposeee. It can be mediated for
instance in loop diagrams with super-symmetric particles (see
figure~\ref{FigMu3eSUSY}), or at tree-level for example via a Z$^\prime$ (see
figure~\ref{FigMu3eZprime}). The current limit set by the SINDRUM experiment
is at $\BR<\num{1.0e-12}$ at $\SI{90}{\percent}$ confidence
level~\cite{Bellgardt:1987du}. 
The upcoming Mu3e experiment plans for a sensitivity of about one signal event
in $\num{e15}$ muon decays in the first phase of the experiment, and  ultimately
one in $\num{e16}$ muon decays in the second phase, thus improving on the
existing limit by four orders of magnitude~\cite{Blondel:2013ia}. 
%% In the following, the experiment in its first phase is presented.

%% The upcoming Mu3e experiment plans for a
%% sensitivity of ultimately one in $\num{e16}$ muon decays, improving on the
%% existing limit by four orders of magnitude~\cite{Blondel:2013ia}. 

%% neutrino oscillations, therefore lepton falvour is violated
%% no observation in the sector of charged leptons so far. but this would be a
%% clear sign for new physics, as by solely incorporating neutrino mixing into
%% the Standard Model, branching fractions of below $\num{e-54}$ are expected.
%% %% observation of \mueee\ would be a clear sign for New Physics \\
%% %% the expectation on the branching ratio for sm plus neutrino mixing is small,
%% %% below 10e-54\\
%% one channel is mu3e. we plan for a sensitivity of ultimately 1 in 1e16,
%% improving on the existing limit by the sindrum experiment by four orders of magnitude\\
%% complementarity of mu3e to other channels

\section{Signal and Background}
The signature of the signal decay \muposeee\ is defined by two positrons and one
electron that emerge from a common vertex and appear coincidently in time. As muon
decays at rest are observed, the energies of the three decay particles sum up
to the muon rest mass and the sum of the momenta vanishes. \\
There are two types of background to \mueee\ searches. One is combinatorial
background arising from the observation of positrons and electrons from
multiple muon decays at a time, e.\,g.\,from the dominant Michel decay of the
muon in combination with an electron-positron pair from Bhabha scattering or
photon conversion. This background can be suppressed via constraints on the
relative timing of the particles, the vertex and momentum. \\
Another type is irreducible background via the radiative muon decay with
internal conversion \muposeeenunu. It can be distinguished from the
signal decay only by the missing energy resulting from the undetected pair of
neutrinos. Thus, the momentum resolution of the detector ultimately limits the
sensitivity of the experiment.

\section{The Mu3e Experiment}
%% The Mu3e experiment aims to search for the decay \mueee\ with a sensitivity of
%% one in $\num{e15}$ muon decays in the first phase of the experiment, and one
%% in $\num{e16}$ in the second phase. In the following, the experiment in its
%% first phase is presented.\\
The aimed at sensitivity level defines the demands on the Mu3e experiment.
%% Reaching the sensitivity goal poses challenges to the Mu3e experiment. 
On the one hand, a large number of muon decays needs to be observed. This
requires not only a high muon stopping rate but also a detector and data
acquisition that is capable to cope with such high rates. 
%% The achievable sensitivity of the experiment is on the one hand limited by the
%% number of observed muon decays. Therefore, high muon rates are required. 
On the other hand, a very good vertex and time resolution and an excellent
momentum resolution are required in order to operate free of background.
%% current limit by sindrum is 1e.12 at 90 percent cl\\
%% mu3e aims at one in e15 in physe 1 and one in e16 in phays 2, improving the
%% sindrum limit ultimately by four orders of magnitude\\
%% this needs high muon stopping rates, 1e8 mu per s in phase\,I, more than e9 in
%% phase 2\\
%% mu3e will operate background-free, a very good vertex and time resolution and
%% an excellent momentum resolution are necessary\\
\subsection{Muon Beam}
%% At the Paul Scherrer Institute, secondary beams of muons are produced by an
%% intense proton beam impinging on a carbon target. 
The Paul Scherrer Institute hosts the world's most intense proton
beam. Secondary continuous muon beams are produced by the protons impinging on
a carbon target. 
Mu3e will be located at the $\uppi$E5 beamline where subsurface muons of
$\SI{28}{\mega\electronvolt\per\text{c}}$ are available. 
The same beamline is also used by the MEGII experiment.
The compact muon beam line (CMBL), that is already installed in the
experimental area of Mu3e, allows to operate both experiments alternately. 
Measurements at the CMBL have shown that the required rates of
$\num{e8}\si{\muon\per\second}$ can be provided.\\
%% The MEGII experiment is also using the $\uppi$E5 beamline.\\
The Paul Scherrer Institute is currently investigating on the high intensity
muon beamline with the aim to provide even higher muon beam rates. This would
enable Mu3e to reach the final sensitivity goal in phase\,II.
%% muons are form the paul-scherrer institute where seondary beams of muons are
%% produced from an intense proton beam impinging on a carbon target.\\
%% e8 of subsurfce muons of 28 MeV over c are available at the existing piE5
%% beamline\\
%% higher rates are under investigation \\
%% the piE5 beamline has to be shared with meg. the compact muon beamline -
%% dedicated beam line for mu3e - is already installed and operational\\
\subsection{Detector Concept}
\begin{figure}[tbp]
  \centering
  \includegraphics[width=\textwidth]{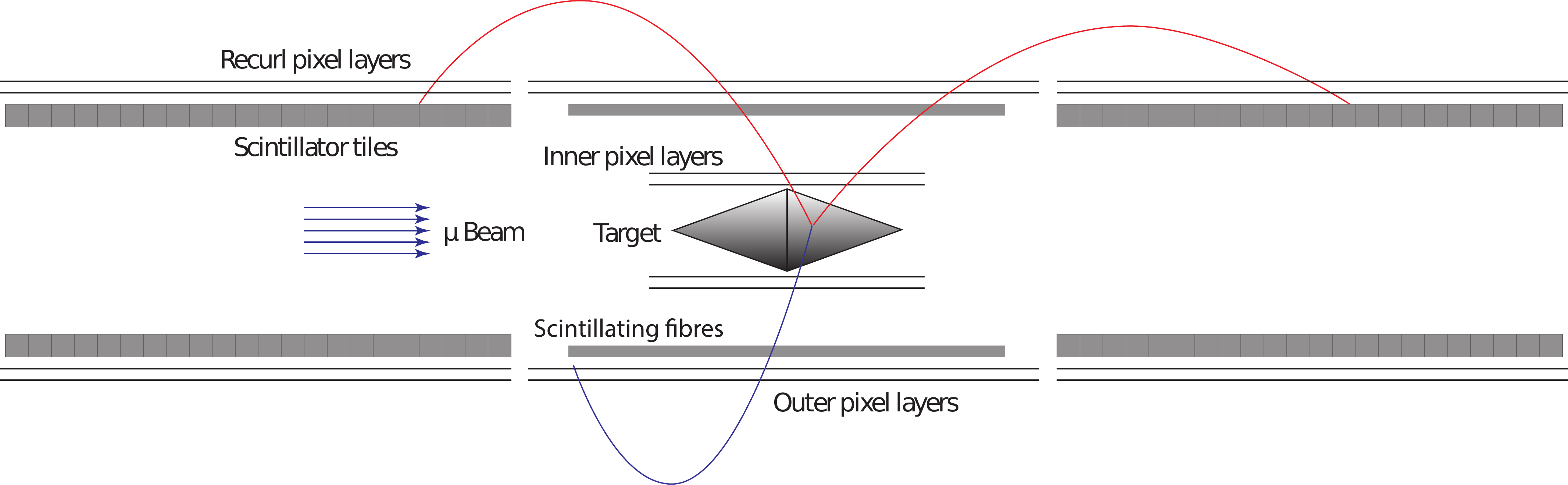}
  \caption{The Mu3e detector in the first phase of the experiment. }
          \label{FigMu3eDetector}
\end{figure}

The geometry of the Mu3e detector is optimized for precise momentum
measurements of the decay electrons and positrons. With a low momentum of
maximally \SI{53}{\mega\electronvolt\per\text{c}}, the momentum resolution is
dominated by multiple Coulomb scattering. Therefore, the material amount in
the active detector volume needs to be kept at a minimum. \\
The momentum is
measured via the bending radius of the particles in a magnetic field. Hence,
the momentum resolution improves with the lever arm between two position
measurements. The optimum momentum resolution is achieved after about a half
%%turn, and this has driven the detector design of Mu3e. \\
turn because at this point uncertainties caused by multiple scattering cancel
to first order. For this reason, the Mu3e experiment relies on measuring
\emph{recurling} tracks. 
When a particle has passed the outer detector layer, it will not cross further
material and following a helical trajectory it will eventually hit the outer
layers again. Between those two measurements, it has performed about a half
turn. \\
%% This means, the trajectory is not only measured of the
%% outgoing but also of the ingoing particles. \\
The geometry of the Mu3e detector is shown in figure~\ref{FigMu3eDetector}. It
has the shape of an elongated tube and is placed in a solenoidal magnetic
field of $\SI{1}{\tesla}$. 
In order to increase the acceptance for recurling tracks, so-called recurl
stations are installed both upstream and downstream of the central detector
part. 
The incoming muons are stopped on a hollow double
cone target made of $\SI{80}{\micro\metre}$ Mylar foil. The shape is chosen to
spread the muon decays over a large surface. \\
The detector consists of a
tracking detector made from thin silicon pixel sensors and a timing detector
system built from scintillating fibres and tiles. 
The thinner fibres are placed in the central detector, the thicker tiles in
the recurl stations as at this point the material amount is no longer crucial.\\
The detector has a total length of about
$\SI{110}{\centi\metre}$ and a diameter of $\SI{18}{\centi\metre}$. 
\subsubsection{Pixel Tracking Detector}
For the tracking detector, pixel sensors are chosen as these comprise not only
an efficient and precise tracking but can also be produced very thin in the
particular technology chosen for Mu3e. \\
%% Momentum resolution is the key to suppress backgrounds. in mu3e multiple
%% coulomb scattering is the dominating uncertainty as the decay electrons have
%% low momenta of 53 MeV at maximum.
%% multiple c scattering can be reduced by minimizing the matieral amaount in the
%% detector.
%% the resolution can be improved by measuring at a large lever arm. if the
%% electron performs about a half turn between two measurements, the uncertainty
%% induced by mcs vanishes to first order. this is the optimal configuration for
%% momentum measurements. \\
%% the mu3e detector consists of a trackind detector made from si pixel sensors
%% and a timing detector system built from scintillating fibres and tiles. the
%% detector has the shape of an elongated tube with a length of 110cm and a
%% diameter of 18cm. It is placed in a solenoidal magnetic field of 1T, so the
%% momentum can be measured by the bending radius. the low momentum muon beam is
%% stopped o a hollow double cone target made from 80 mum of mylar foil on
%% average. the shape of the target is chosen to distribute the muon decay
%% verteices over a large surface to facilitate the suppression of combinatorial
%% background. 
The target is surrounded by two layers of pixels sensors which allows for a
precise determination of the decay vertex. Two more layers of pixel sensors
are placed at a larger radius enabling the momentum measurement.
%% Leaving the outer pixel layers, the decay
%% particles will not cross further material and, as they move on a helical
%% trajectory, they will eventually hit the outer layers again. Between those two
%% measurements in the outer layers, the lever arm is particularly large.
%% shorlty underneath the outer pixel layers there will be a timing detector made of
%% scintillaing fibres. 
%% In order to increase the acceptance for these recurling tracks and thus
%% to profit from the smallest possible momentum resolution, so-called recurl
%% stations are installed both upstream and downstream of the central detector
%% part. 
The recurl stations are equipped as well with a double-layer of pixel sensors
with the same radii as the outer pixel layers in the central station. \\
%% the aim of the pixel tracke is to measure low momentum electrons with high
%% precision and thus it is required to consist only of very little
%% material. Mu3e relies on thin silicon pixels sensors. these will be read out
%% and powered over flexible printed circuit boards. a support structure made
%% from kapton holds the layers in place. Thus, one layer of pixel sensors
%% accounts only for 1.16 permille of a radiation length. \\
\begin{figure}[tbp]
  \centering
  \includegraphics[width=0.4\textwidth]{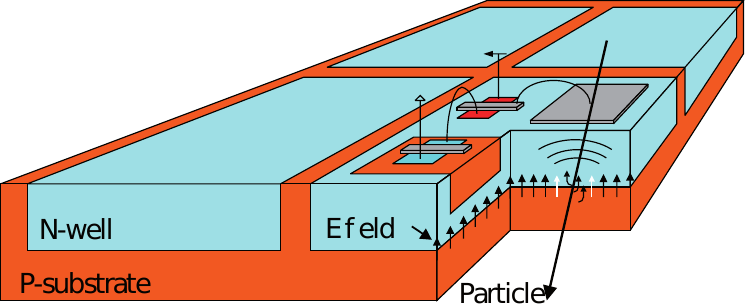}
  \caption{Schematic of a HV-MAPS~\cite{Peric:2007zz}. }
          \label{FigHVMAPS}
\end{figure}
Mu3e deploys High Voltage Monolithic Active Pixel
Sensors~\cite{Peric:2007zz}. The individual pixels are implemented as deep N-wells
in a p-doped substrate (see figure~\ref{FigHVMAPS}). By applying a reverse
bias voltage of about \SI{80}{\volt}, a depletion zone of \num{10}\ to
\SI{20}{\micro\metre}\ is formed. The electric field in the depletion zone
is strong, so that charge carriers created by traversing ionising radiation can
be collected via drift. As particle detection is limited to a thin volume
close to the surface of the sensor, these sensors can be made very
thin. 
For the final chip, a thickness of \SI{50}{\micro\metre}\ is envisaged.
%% Current prototypes are only \SI{50}{\micro\metre}\ thin. The same
%% thickness is envisaged for the final chip.

\begin{figure}[tbp]
  \centering
  \subfloat[Efficiency.]{\includegraphics[height=0.15\textheight]{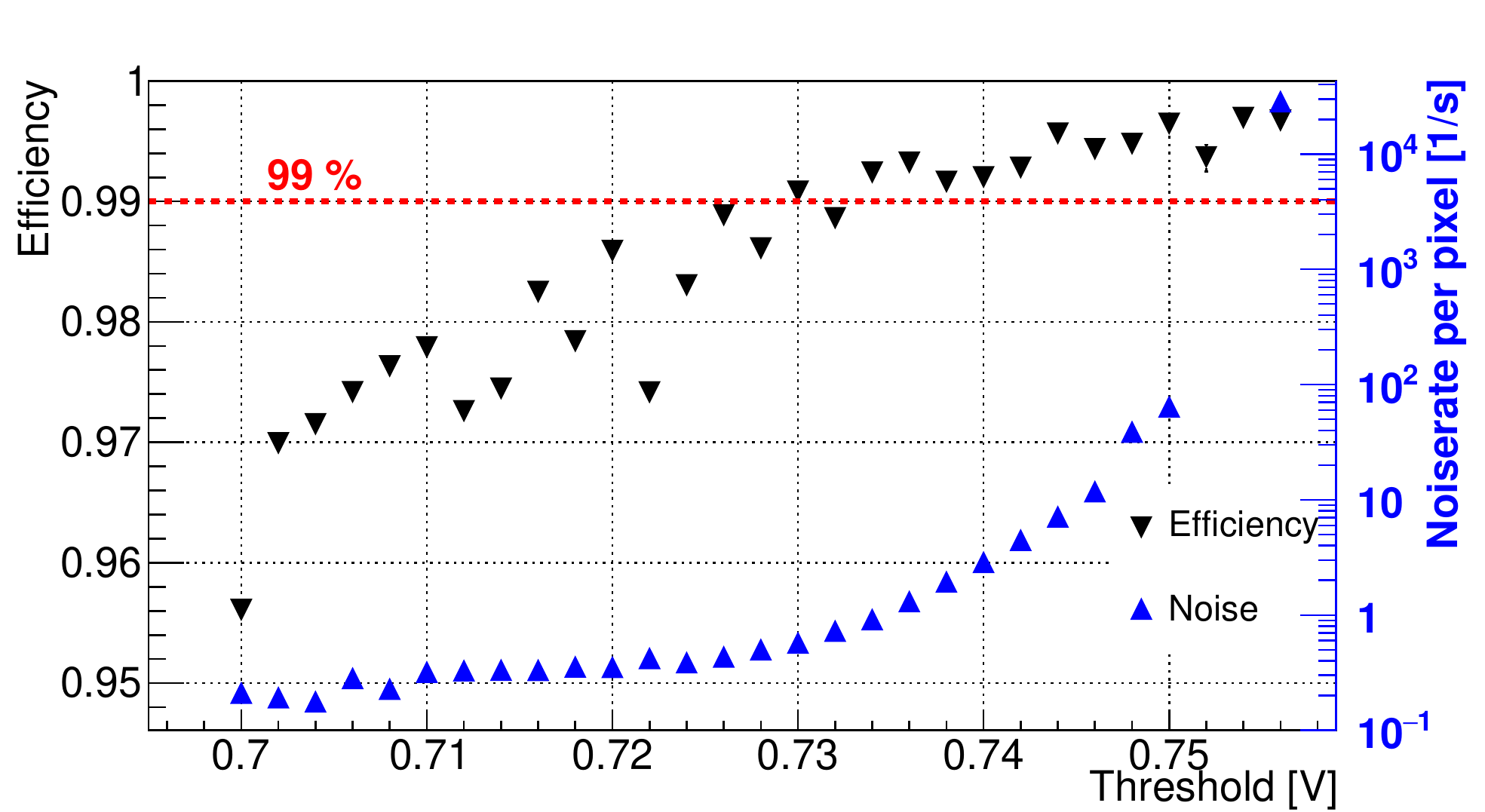}\label{FigMupixEff}}\quad
  \subfloat[Timing resolution.]{\includegraphics[height=0.15\textheight]{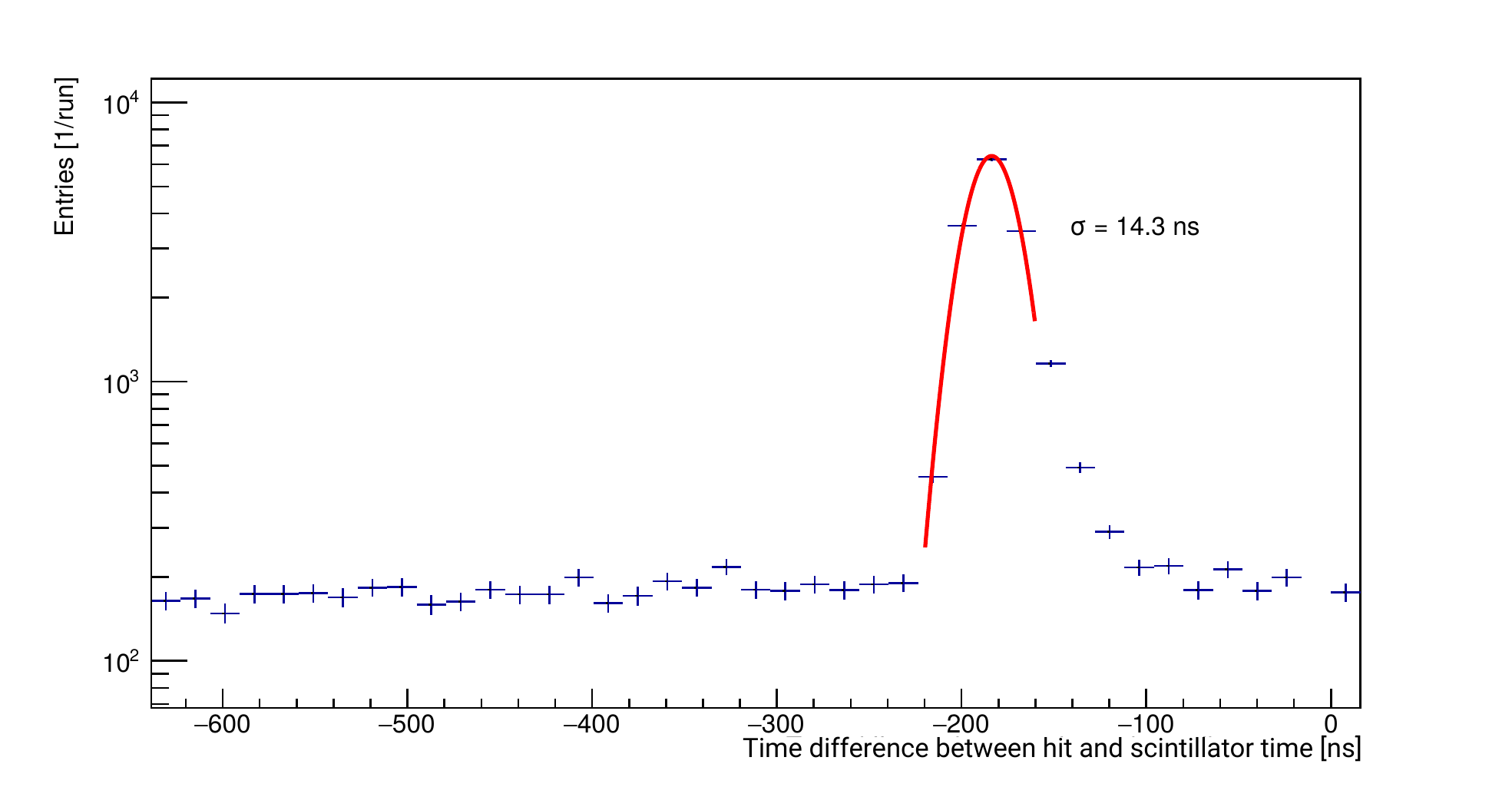}\label{FigMupixTime}} 
  \caption{Efficiency and timing resolution obtained with the MuPix7 prototype
    in testbeam measurements. }
          \label{FigMupixResults}
\end{figure}

Furthermore, it is possible to implement transistors within the
pixel and thus build readout electronics directly on the sensor chip. Signal
amplification and shaping is performed in the pixel itself, whereas the
digitisation happens in the periphery, a small part at the bottom edge of the
sensor. The sensor has digital, zero-suppressed data output at a fast serial
link of $\SI{1.25}{\giga\bit\per\second}$.\\
For the final chip, a pixel size of \num{80x80}\,\si{\micro\metre\squared}\ is
planned with an active area of \num{2x2}\,\si{\centi\metre\squared}\ per sensor.\\
The development of the sensors is currently in the prototyping stage. The
MuPix7 prototype has an active area of
\num{2.9x3.2}\,\si{\milli\metre\squared}\ with a pixel size of
\num{103x80}\,\si{\micro\metre\squared}. Despite the small size, this prototype
comprises already all the functionalities required in the final chip. The
MuPix7 has been extensively tested with test beam, yielding for instance an
efficiency well above \SI{99}{\percent}\ and a timing resolution smaller
than \SI{20}{\nano\second}~\cite{Augustin:2016hzx} (see
figure~\ref{FigMupixResults}). Prototypes with a thickness of
\SI{50}{\micro\metre}\ are fully functional and perform equally well as
thicker prototypes. \\
The latest prototype MuPix8 is the first large scale MuPix sensor with an
active area of \num{2x1}\,\si{\centi\metre\squared}\ and  a pixel size of
\num{80x81}\,\si{\micro\metre\squared}. 
%% , allowing
%% to test the performance when the analog signal is send over a long line from
%% the pixel matrix to the periphery. 
In order to improve the timing resolution,
the MuPix8 incorporates signal amplitude measurements allowing to correct for
time-walk. 
%% the MuPix8 provides information about the analogue pulse for instance by the
%% output of a time-over-threshold. 
The MuPix8 has been submitted on two different substrates: One with a
resistivity of \SI{20}{\ohm\centi\metre}, the same as in the case of MuPix7, and one
with a higher resistivity of \SI{80}{\ohm\centi\metre}. With the higher
resistivity, the signal becomes twice as large, and thus the signal to noise
ratio potentially improves. \\
The characterisation of MuPix8 has just started. First measurements have
shown that the chip is operational and indicate an overall good performance. \\
%% !!! TODO: MUPIX8 RESULTS !!!\\
%% say that it is back since august and tested in teh lab\\
A further prototype, MuPix9, has recently been submitted. It is a small scale
prototype comprising circuitries for slow control that
have not yet been implemented in previous prototypes.
In addition, a novel serial-powering scheme is investigated. \\
%% something about the fast lvds link \\
The pixel sensors produce heat with a power of about
\SI{250}{\milli\watt\per\centi\metre\squared}\ and thus need to be cooled. In
order not to add too much material to the detector, this is performed with a
flow of gaseous helium, both globally in the whole detector volume but also
locally through channels integrated in the mechanical support of the pixel
sensors. Finite element simulations predict feasible temperature gradients of
about \SI{70}{\kelvin}. 
The simulation results are verified in tests with a thermal mock-up of the
detector. 
Although high flow velocities of \SI{20}{\metre\per\second}\ are applied
locally, measurements have proven that
flow-induced vibrations stay below \SI{10}{\micro\metre}\ and will thus not
affect the track measurements. 

\subsubsection{Scintillating Timing Detector}
\begin{figure}[tbp]
  \centering
  \subfloat[Prototype with three layers of square multiclad fibres.]{\includegraphics[height=0.2\textheight]{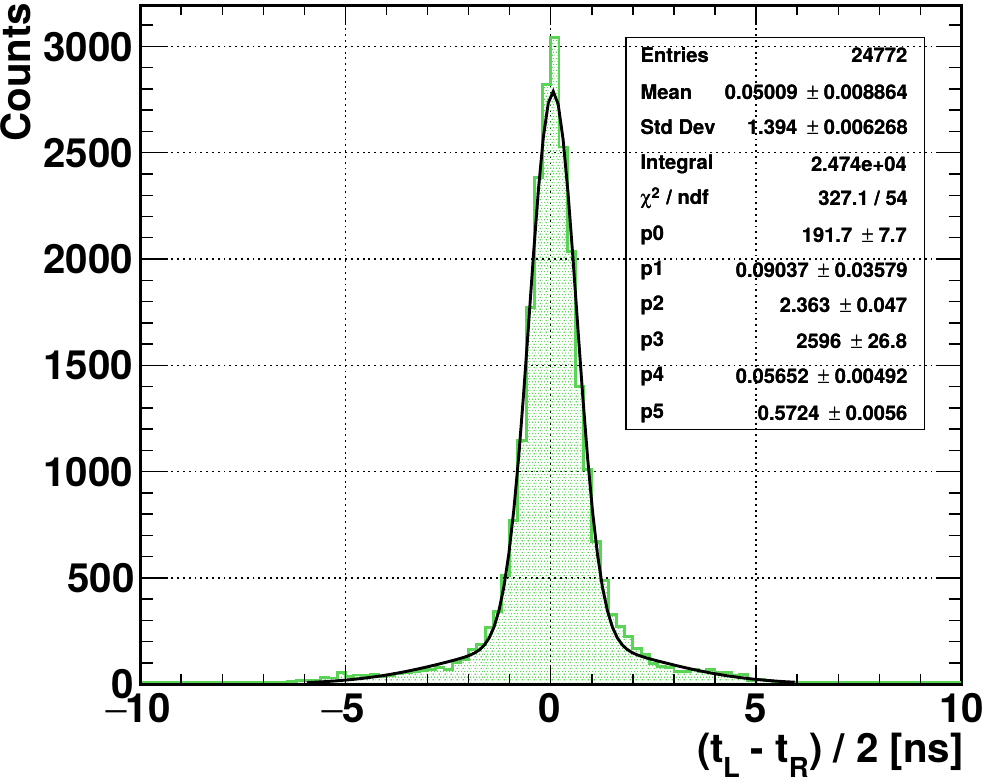}\label{FigFibres}}\quad
  \subfloat[Tile detector prototype. The time difference is shown with and
    without time-walk correction (TWC).]{\includegraphics[height=0.21\textheight]{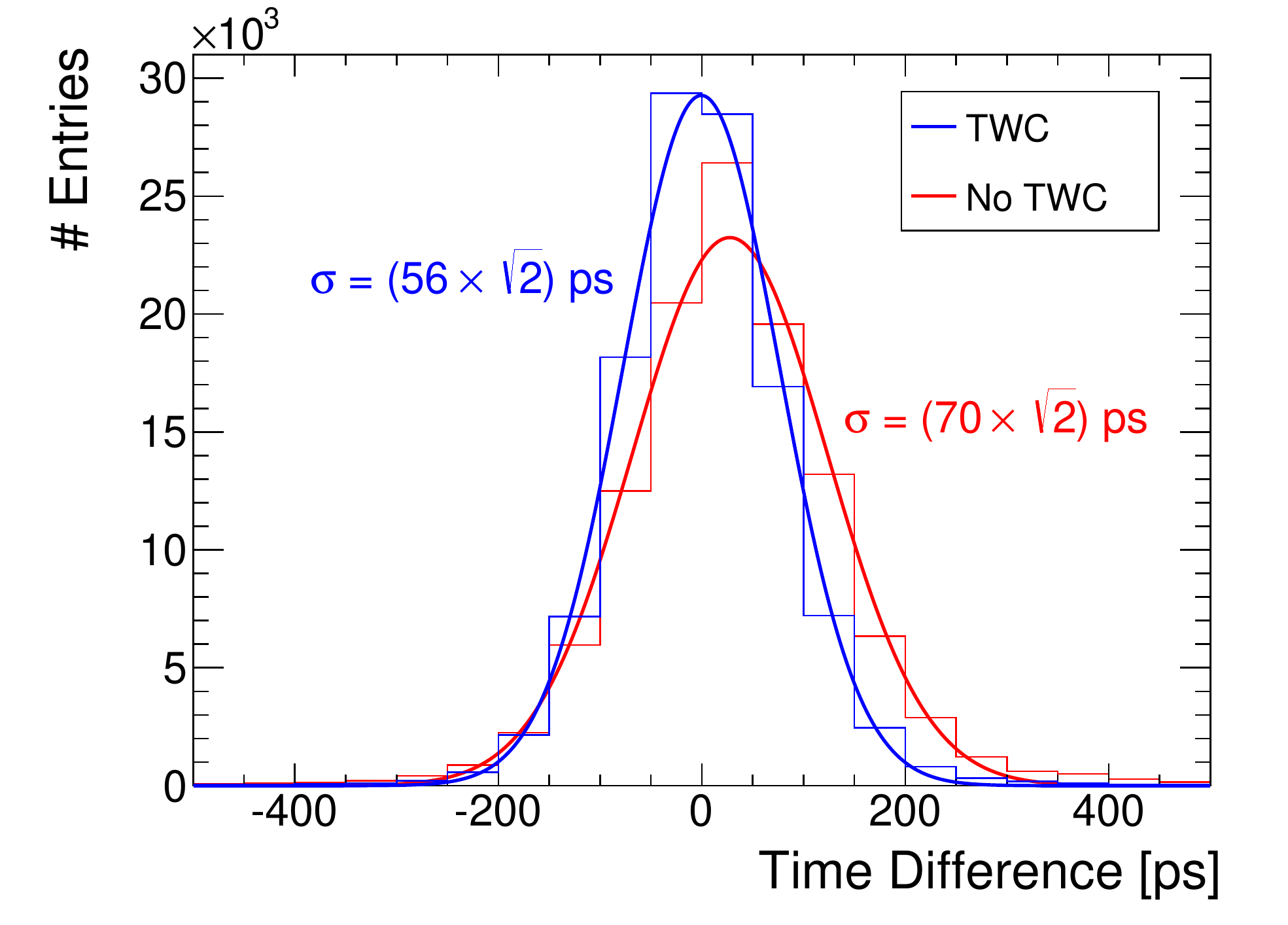}\label{FigTiles}} 
  \caption{Timing resolution of prototypes for the scintillating timing
    detectors measured in test beam. }
          \label{FigScintillators}
\end{figure}

The combinatorial background can be efficiently suppressed by timing
measurements. In the central detector part, a thin timing detector made of
three layers of scintillating fibres is placed directly underneath the outer
pixel layers. Likewise, the recurl stations are equipped with scintillating
tiles. %% underneath the pixel layers.
With the combination of timing information from the fibre and
tile detector, suppression factors of about 100 for backgrounds with two
correlated and one uncorrelated track can be achieved. Backgrounds with three
uncorrelated tracks are even stronger suppressed. \\
%%!!! TODO: SHOW PLOT? !!!\\
The scintillation photons of the fibres are collected at both ends with a
silicon photo-multiplier (SiPM) column array as used by the LHCb
experiment~\cite{LHCbCollaboration:2014tuj}. Currently, fibres with round and
square cross sections are under investigation, and both fulfill the
requirements for the use in Mu3e.
For example, a fibre prototype consisting of three layers of square multiclad
fibres achieves timing resolutions of \SI{572}{\pico\second}\ with an
efficiency above \SI{95}{\percent} (see figure~\ref{FigScintillators}).\\
Each of the scintillating tiles in the recurl stations has an individual
SiPM. In both detector systems, the SiPMs are read out with an custom-designed
time-to-digital converter ASIC called MuTRiG~\cite{Chen:2017qor}.
In the case of the tile detector, the MuTRiG operates in a two-threshold mode
which allows for time-walk compensation.
%% In the case of the tile detector, it is even possible to perform time walk
%% correction on the MuTRiG.
%% These
%% are readout by a custom designed MuTriG. The MuTRiG is a tdc asic designed in
%% the context of the tile detecotr. in the case of tiles, it is capable to perform time walk
%% correction and thus further improve the timing resolution. the tile detector
%% is made of 6.5x6.5x5.0mm3 tiles each with an individual SiPM for photon
%% detection.\\
Prototypes of the tile detector yield time resolutions of
\SI{70}{\pico\second}\ without and \SI{56}{\pico\second}\ with time walk
correction at an efficiency of \SI{99.7}{\percent}. 
%% A fibre prototype
%% consisting of three layers of square multiclad fibres achieves timing
%% resolutions of \SI{572}{\pico\second}\ with an efficiency above
%% \SI{95}{\percent} (see figure~\ref{FigScintillators}).\\

\section{Sensitivity Studies}
\begin{figure}[tbp]
  \centering
  \subfloat[Reconstructed muon mass for signal and background processes. ]{\includegraphics[height=0.23\textheight]{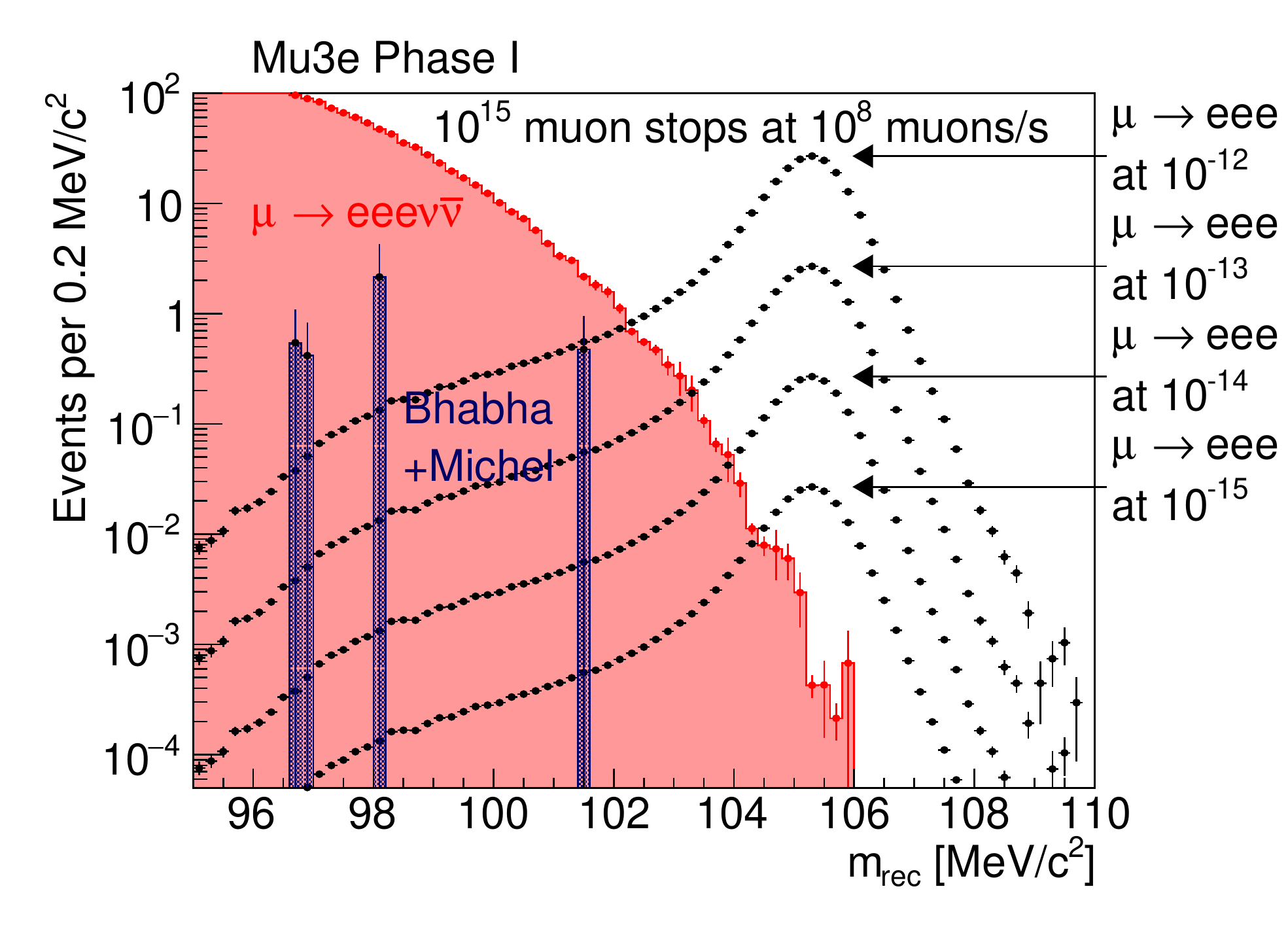}\label{FigMu3eMass}}\quad
  \subfloat[Expected sensitivity to \mueee. ]{\includegraphics[height=0.215\textheight]{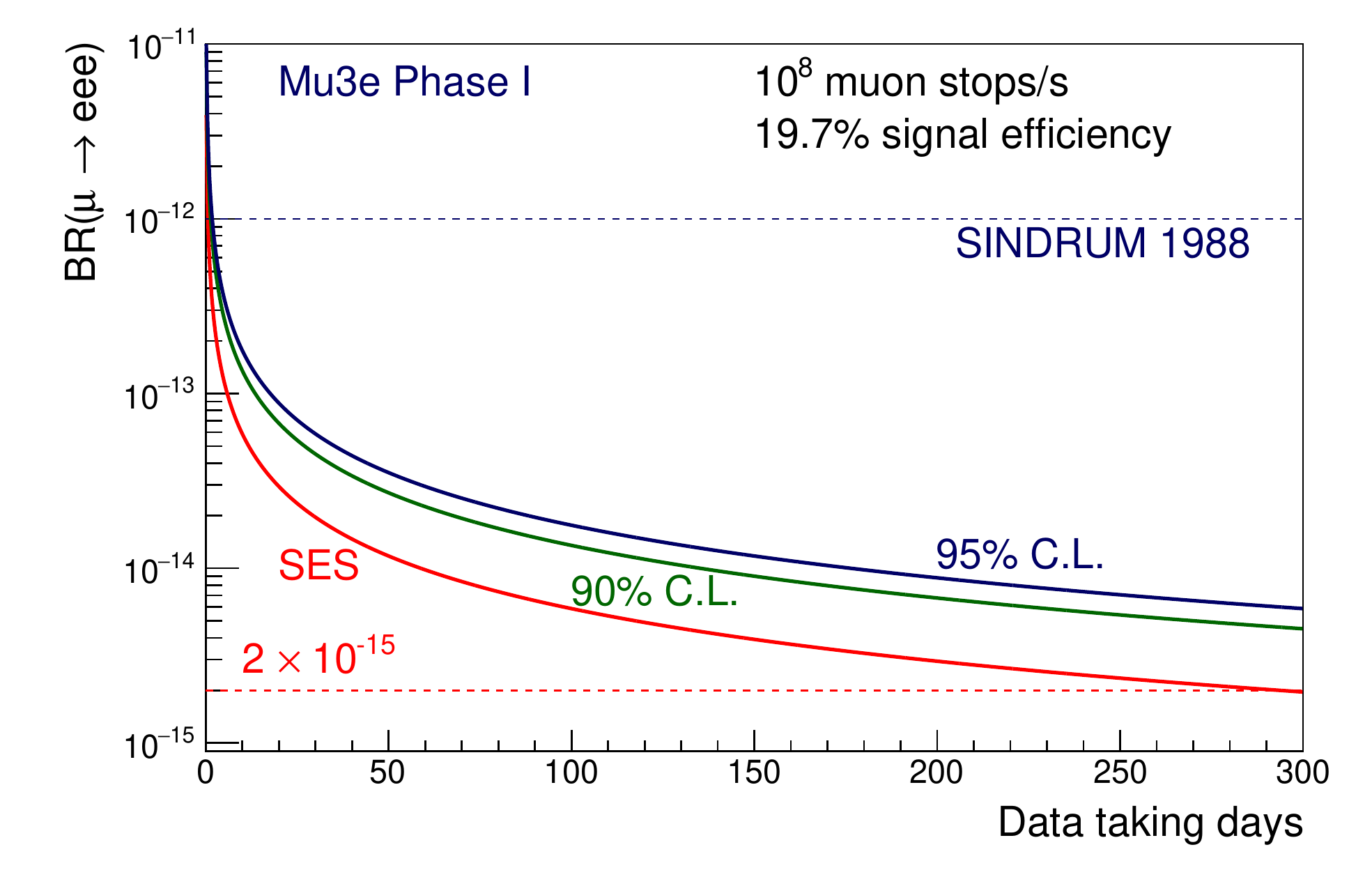}\label{FigMu3eSensitivity}} 
  \caption{Simulation results for the Mu3e experiment in the first phase. }
          \label{FigMu3eSimulation}
\end{figure}
The sensitivity of Mu3e to the decay \mueee\ is estimated using a detailed
Geant4 simulation of the detector. 
In figure~\ref{FigMu3eMass}, the reconstructed mass of the three
electron-system is shown for the potential signal decay at various branching
fractions as well as for background from internal conversion decays and
combinatorial background. The combinatorial background is here represented by
the dominant contribution of Michel decays in combination with a Bhabha
scattering event in which a positron transfers enough momentum to an electron
such that the electron tracks is also visible in the detector.\\
With an expected run time of 300 days of data taking in phase\,I at a muon
stopping rate of \num{e8}\si{\per\second}, the expected single-event
sensitivity is found to be \num{2e-15}, corresponding to a branching fraction
limit of \num{4e-15}\ at \SI{90}{\percent}\ confidence level (see
figure~\ref{FigMu3eSensitivity}). 

\section{Status}
Prototypes of the various subdetectors have successfully proven the
suitability of the chosen technologies for the usage in the Mu3e
experiment. The outcome of the studies of MuPix8 and MuPix9 for instance will
pave the way to the final pixel sensor chip. \\
The collaboration is currently finalizing the detector design and preparing
for construction and commissioning of the experiment. Commissioning is
foreseen for 2019 with potential first physics data taking in 2020.\\
In the first phase of Mu3e, a single event sensitivity of \num{2e-15}\ is
expected which poses a significant improvement on the limit to \mueee\ decays
with respect to previous experiments. The final sensitivity of Mu3e will be
reached in a second phase which is planned to be an upgrade of the phase\,I
detector. The envisaged single event sensitivity of \num{1e-16}\ is however
only feasible when muon stopping rates of about
\num{2e9}\si{\per\second}\ become available. 

%% what to say about the status?
%% mupix8 running. a positive outcome of mupix8 and mupix9 measurements will path
%% the way for a chip suitable for module building
%% magnet in first half of 2019\\
%% phase 2\\
%% finalizing detector design, preparing for construction and commissioning

\acknowledgments
The author acknowledges support by the Research Training Group `Particle
Physics beyond the Standard Model' (GRK\,1940) of the Deutsche Forschungsgemeinschaft.
%% \appendix

\end{document}